\documentclass[a4paper,10pt]{article}
\usepackage{amsthm}
\usepackage{amsfonts}
\usepackage{amssymb}
\usepackage{mathrsfs}
\usepackage[english]{babel}
\usepackage{float}
\usepackage[toc,page]{appendix}
\usepackage[pdftex]{graphicx}
\usepackage{epstopdf}
\usepackage{amsmath}
\usepackage{color}

\numberwithin{equation}{section}

\newtheorem{theorem}{Theorem}[section]

\newtheorem{definition}[theorem]{Definition}

\newtheorem{lemma}[theorem]{Lemma}

\newtheorem{proposition}[theorem]{Proposition}

\begin{document}

\title{Exact recovery of Dirac ensembles from the projection onto spaces of spherical harmonics}
\date{October 2013, revised May 2014}
\author{Tamir Bendory, Shai Dekel and Arie Feuer}
\maketitle

\begin{abstract}

In this work we consider the problem of recovering an ensemble of Diracs on the sphere from its 
projection onto spaces of spherical harmonics. We show that under an appropriate separation condition on 
the unknown locations of the Diracs, the ensemble can be recovered through Total Variation norm minimization. The 
proof of the uniqueness of the solution uses the method of `dual' interpolating polynomials and is based 
on \cite{candes2013towards}, where the theory was developed for trigonometric polynomials. We also show that
in the special case of non-negative ensembles, a sparsity condition is sufficient for exact recovery. 

\end{abstract}



\section{Introduction}

In many cases, images and signals are observed on spherical manifolds. Typical examples are astrophysics (e.g. \cite{komatsu2011seven}), topography \cite{audet2011directional} and gravity fields sensing \cite{klosko1982spherical}. Further example is spherical microphone arrays,  used for spatial beam forming \cite{meyer2001beamforming} and sound recording   \cite{meyer2003spherical}. 

A key tool for the analysis of signals on the sphere is spherical harmonics analysis, discussed in detail later on. For instance, the spherical microphone array was analyzed in terms of spherical harmonics in \cite{rafaely2005analysis}.  Additionally, spherical harmonics have been extensively used for various applications in computer graphics, such as modeling of volumetric scattering effects, bidirectional reflectance distribution function, and atmospheric scattering (for more graphical applications, see \cite{sloan2008stupid} and the references therein). Spherical harmonics are also used in medical imaging \cite{taguchi2001cone}, optical tomography \cite{arridge1999optical}, several applications in physics such as solving potential problem in electrostatics \cite{macrobert1928spherical} and the central potential  Schr\"odinger equation in quantum mechanics \cite{cohen2006quantum}. Additional applications of spherical harmonics are sampling on the sphere  \cite{mcewen2011novel,ben2012generalized} and more recently, compressed sensing \cite{alem2012sparse} and sparse 
recovery \cite{Rauhut_sparse,mcewen2012sparse}. In some sense, our work relates to these latter fields. 

Let $\mathcal{H}_n({\mathbb{S}^{d-1}})$ denote the space of homogeneous spherical harmonics of degree $n$, which is the restriction to the unit sphere of the homogeneous harmonic polynomials of degree $n$ in $\mathbb{R}^d$ \cite{atkinson_spherical}. Each subspace $\mathcal{H}_n({\mathbb{S}^{d-1}})$ is of dimension
\begin{equation*}
a_{n,d}:=\frac{(2n+d-2)(n+d-3)!}{n!(d-2)!},\quad  n\in\mathbb{N},d\geq 2.
\end{equation*}
Also, recall that $L_2(\mathbb{S}^{d-1})=\oplus_{n=0}^\infty\mathcal{H}_n({\mathbb{S}^{d-1}})$. 
Thus, if $\{ Y_{n,j} \}$, $j=1,..., a_{n,d}$, is an orthonormal basis of $\mathcal{H}_n({\mathbb{S}^{d-1}})$, then 
$f\in L_2(\mathbb{S}^{d-1})$ can be expanded as $f=\sum_{n=0}^\infty f_n$, where 
\begin{align*}
f_n&=\sum_{j=1}^{a_{n,d}} \langle f,Y_{n,j}\rangle Y_{n,j}.
\end{align*}
Using the \emph{Addition Formula} \cite{atkinson_spherical}, one can write the kernel of the projection onto 
$\mathcal{H}_n({\mathbb{S}^{d-1}})$ as
\begin{equation} \label{eq:addition_formula}
\mathcal{P}_{n,d}(\zeta\cdot\eta)=\sum_{j=1}^{a_{n,d}}Y_{n,j}(\zeta)\overline{Y_{n,j}(\eta)}=
\frac{a_{n,d}}{|\mathbb{S}^{d-1}|}P_{n,d}(\zeta\cdot\eta), \qquad \zeta,\eta \in \mathbb{S}^{d-1}, 
\end{equation}
where $P_{n,d}$ is univariate ultraspherical Gegenbauer  polynomial of order $d$ and degree $n$. 
Thus, the projection kernel onto the space $V_N:=\oplus_{n=0}^N \mathcal{H}_n({\mathbb{S}^{d-1}})$ is given by
\begin{equation} \label{ortho-proj}
K_{N}(\zeta\cdot\eta):=\sum_{n=0}^{N}\mathcal{P}_{n,d}(\zeta\cdot\eta).
\end{equation}

\noindent In this work we consider the Dirac ensemble 
\begin{equation} \label{eq:signal}
f=\sum_m c_m\delta_{\mathbf{\xi}_m},
\end{equation}
where $\delta_x$ is a Dirac measure, $c_m\in\mathbb{R}$ are real weights, and $\xi_m\in \Xi\subset 
\mathbb{S}^{d-1}$ are distinct locations on the sphere. We recall the following definition 

\begin{definition}
Let $\mathcal{B}(A)$ be the Borel $\sigma$-Algebra on a compact space $A$, and denote by $\mathcal{M}(A)$ the associated space of real Borel measures. The Total Variation of a real Borel measure $v\in \mathcal{M}(A)$ over a set $B\in\mathcal{B}(A)$ is defined by 
\begin{equation*}
\vert v\vert (B)=\sup\sum_k\vert v(B_k)\vert,
\end{equation*}
where the supremum is taken over all partitions of $B$ into a finite number of disjoint measurable subsets. The total variation $\vert v \vert$ is a non-negative measure on $\mathcal{B}(A)$, and the Total Variation (TV) norm of $v$ is defined as
\begin{equation*}
\|v\|_{TV}=\vert v\vert (A).
\end{equation*} 
\end{definition}
\noindent For a measure of the form of (\ref{eq:signal}), it is easy to see that
\begin{equation} 
\|f\|_{TV}=\sum_m\vert c_m\vert.
\end{equation}
In this paper we assume that the only information we have on the signal $f$ is its `orthogonal projection' onto $V_{N}$, i.e, 
\begin{equation}
y_{n,j}:=\langle f,Y_{n,j} \rangle=\sum_m c_m Y_{n,j}(\xi_m),	\quad 0 \le n \le N, \quad 1 \le j \le a_{n,d}.
\label{eq:fN}
\end{equation}
To ensure exact recovery of the Dirac ensemble from its projection onto $V_N$, we impose a separation
condition as in \cite{candes2013towards} for the case of trigonometric polynomials 
and \cite{bendory2013exact} for the case of algebraic polynomials over $[-1,1]$. 
To this end, recall that the distance on the sphere between any two points $\xi_1,\xi_2\in\mathbb{S}^{d-1}$ is given by
\begin{equation} \label{def-dist}
d(\xi_1,\xi_2)=\arccos\left(\xi_1\cdot\xi_2\right).
\end{equation}

\begin{definition} \label{def:separation}
A set of points $\Xi\subset \mathbb{S}^{d-1}$ is said to satisfy the minimal separation condition for
(sufficiently large) $N$ if   
\begin{equation}
\Delta:=\min_{\xi_i,\xi_j\in\Xi, \xi_i\neq\xi_j}d\left(\xi_i,\xi_j\right)\geq \frac{\nu}{N},  
\end{equation}
where $\nu$ is a fixed constant that does not depend on $N$.
\end{definition}

\noindent The main theorem of this paper concerns exact recovery in the case $d=3$, i.e. the sphere $\mathbb{S}^2$

\begin{theorem}
Let $\Xi=\{\xi_m\}$ be the support of a signed measure of the form (\ref{eq:signal}). Let $\{Y_{n,j}\}_{n=0}^N$ be any spherical harmonics basis for $V_N(\mathbb{S}^2)$ and let $y_{n,j}=\langle f,Y_{n,j}\rangle$, $0\le n \le N$, $ 1 \le j \le a_{n,3}$. 
If $\Xi$ satisfies the separation condition of Definition \ref{def:separation}, then $f$ is the unique solution of
\begin{equation}
\begin{split}
\min_{g\in \mathcal{M}(\mathbb{S}^2)}\|g\|_{TV} \quad &\mbox{subject to} \quad \langle g,Y_{n,j}\rangle = y_{n,j},  \\
n=0,...,N, &\quad j=1,...,a_{n,3},
\end{split}
\end{equation}
where $\mathcal{M}(\mathbb{S}^2)$ is the space of signed Borel measures on $\mathbb{S}^2$.
\label{th:main}
\end{theorem}

Observe that for applications, Theorem \ref{th:main} is stronger than needed. Indeed, since the form of (the unknown) $f$ 
is known, one may perform TV minimization over the smaller subspace of Dirac superpositions over the sphere. Practical 
numerical algorithms that leverage on this result are presented in \cite{bendory2014practical}. 
Also, we strongly believe that this result holds in higher dimensions and indeed significant parts of the proof can be 
easily generalized to any dimension. However, there are certain technical challenges (see Section \ref{subsection-inversion}) which we hope to overcome in future work.

The outline of the paper is as follows. In Section \ref{sec:dual} we recall the dual problem of interpolating polynomials.
In Section \ref{sec:loc} we provide details on the essential ingredient of the dual polynomial construction, which is 
a well-localized polynomial kernel. In Section \ref{sec:construct} we carry out the actual construction of the interpolating polynomial. In Section \ref{sec:non-negative} we review the simpler case of signals with non-negative coefficients,
where the separation condition can be replaced by a significantly weaker assumption of sparsity. i.e. that the
number of Diracs is $\le N$. 

Finally, we point out that the main result of the paper is of qualitative nature in the following sense. 
Throughout the proofs we will have for some $k \ge 3$, elements of the type $c_k / \nu^{k-1}$, where $c_k$ 
are absolute constants that depend only on $k$, but change from estimate to estimate and $\nu$ is the constant 
from Definition \ref{def:separation}. Once all estimates are done, $\nu$ is selected to be sufficiently large 
so that $c_k / \nu^{k-1}$ and  similar quantities are sufficiently small. In this paper, we do not deal
with the problem of the sharpness of the constant $\nu$.

\section{The dual problem of polynomial interpolation} \label{sec:dual}

The proof of Theorem \ref{th:main} can be reduced to a problem in polynomial interpolation. This result in its general form is given in  \cite{bendory2013exact} (see also \cite{candes2013towards,de2012exact}). For completeness, we provide here the proof for the case of real coefficients

\begin{theorem}
Let $f=\sum_m c_m\delta_{\xi_m} , c_m\in\mathbb{R}$, where $\Xi:=\{\xi_m\}\subseteq A$, and $A$ is a compact manifold in 
$\mathbb{R}^n$. Let $\Pi_D$ be a linear space of continuous functions of dimension $D$ in $A$. For any basis $\{P_k\}_{k=1}^D$ of $\Pi_D$, let $y_k = \langle f,P_k\rangle$ for all $1\leq k\leq D$. If for any set $\{ u_m \}$, $u_m\in\mathbb{R}$, with $\vert u_m\vert=1$, there exists $q\in \Pi_D$ such that
\begin{align*}
q(\xi_m)&=u_m \,,\, \forall \xi_m\in \Xi  ,\label{eq:dual1} \\
\vert q(\xi)\vert&<1 \,,\, \forall \xi\in A\backslash \Xi, 
\end{align*}
then $f$ is the unique real Borel measure satisfying 
\begin{equation} \label{eq:dual1}
\min_{g\in\mathcal{M}(A) }\|g\|_{TV} \quad \mbox{subject to} \quad y_k = \langle g,P_k\rangle  \,,\, 1\leq k\leq D.
\end{equation}
\label{th:duality}
\begin{proof}
Let $g$ be a solution of (\ref{eq:dual1}), and define $g=f+h$. 
The difference measure $h$ can be decomposed relative to $\vert f\vert$ as 
\begin{equation*}
h=h_{\Xi}+h_{\Xi^C},
\end{equation*}
where $h_\Xi$ is concentrated in $\Xi$, and $h_{\Xi^C}$ is concentrated in $\Xi^C$ (the complementary of $\Xi$). 
Note that if $h_\Xi=0$, than $h_{\Xi^C}=0$ also, otherwise $\|g\|_{TV} > \|f\|_{TV}$, which is a contradiction. 
Thus, in such as case, $h=0$ and $f$ is the unique minimizer of (\ref{eq:dual1}). Performing a polar decomposition of $h_\Xi \ne 0$ yields
\begin{equation*}
h_\Xi=\vert h_\Xi\vert sgn(h_\Xi)(\xi),
\end{equation*}
where $sgn(h_\Xi)$ is a function on $A$ with values $\{-1,1\}$ (see e.g. \cite{rudin}). 
By assumption, there exists $q\in \Pi_D$ obeying 
\begin{align}
q(\xi_m)&= sgn(h_\Xi)(\xi_m) \,,\, \forall \xi_m\in \Xi ,\label{eq:q_xm}\\
\vert q(\xi)\vert&<1 \,,\, \forall \xi\in A\backslash \Xi . \label{eq:q_x}
\end{align}
Also by assumption $\langle g,P_k\rangle = \langle f,P_k\rangle$, for $1\leq k\leq D$, and so
\begin{equation}
\langle q,h\rangle = 0.
\label{eq:in_pro}
\end{equation}
Equation (\ref{eq:in_pro}), with the polar decomposition of $h_\Xi$ and (\ref{eq:q_xm})  imply
\begin{equation*}
0=\langle q,h_\Xi \rangle+\langle q,h_{\Xi^C}\rangle=\|h_\Xi \|_{TV}+\langle q,h_{\Xi^C}\rangle.
\end{equation*}
If $h_{\Xi^C}=0$, then $\|h_\Xi\|_{TV}=0$, and $h=0$. Alternatively, if $h_{\Xi^C}\neq 0$ ,we conclude by property (\ref{eq:q_x}) that 
\begin{equation*}
\vert \langle q,h_{\Xi^C}\rangle \vert  < \|h_{\Xi^C}\|_{TV}.
\end{equation*}
Thus, 
\begin{equation}
\|h_{\Xi^C}\|_{TV}>\|h_{\Xi}\|_{TV}.
\label{eq:nsp}
\end{equation}
As a result of (\ref{eq:nsp}), we get 
\begin{equation*}
\begin{split}
\|f\|_{TV}&\geq \|f+h\|_{TV}=\|f+h_\Xi \|_{TV}+\|h_{\Xi^C}\|_{TV} \\
&\geq \|f\|_{TV}-\|h_\Xi \|_{TV}+\|h_{\Xi^C}\|_{TV}>\|f\|_{TV} ,
\end{split}
\end{equation*}
which is a contradiction. 
Therefore, $h=0$, which implies that $f$ is the unique solution of (\ref{eq:dual1}). 
\end{proof}
\end{theorem}

In the Figure below, we see an example of an interpolating spherical harmonic polynomial 
$q:\mathbb{S}^2 \rightarrow [0,1]$ where $N=50$. The heat map shows dark red at points $\xi_m \in \Xi$, 
where $q(\xi)=1$ and blue in regions where $q$ is close to zero.

\begin{figure}[ht]
\begin{center}
\includegraphics[width=0.5\textwidth,height=0.5\textwidth]{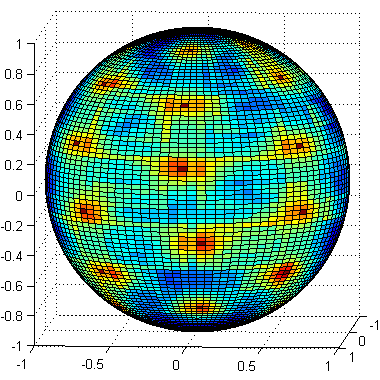}
\end{center}
\caption{ An interpolating polynomial on the sphere.}
\label{fig:construction}
\end{figure}

\section{Spherical Harmonics localization} \label{sec:loc}

It is well known that the orthogonal projection kernel $K_{N}$ given by (\ref{ortho-proj}) 
does not have good localization. Instead, we follow \cite{petrushev_decomposition} and for $d=3$ define the kernel
\begin{equation}
\widetilde{F}_{N}(\zeta\cdot\eta):=\sum_{n=0}^\infty\rho(n/N)\mathcal{P}_{n,3}(\zeta\cdot\eta),
\end{equation}
where $\rho\in C^\infty[0,\infty)$ is a smooth non-negative univariate function, satisfying 
\begin{equation} \label{def-rho}
\rho(t)=\begin{cases} 1, &\quad t\in[0,1/2],  \\  \leq 1, &\quad t\in[1/2,1], \\ 0, &\quad \mbox{otherwise.}\end{cases} 
\end{equation} 

\noindent We emphasize that $\widetilde{F}_{N}(\cdot)$ can be regarded as a superposition of Gegenbauer polynomials of degree
$\le N$ and hence also a univariate algebraic polynomial of degree $N$. Let us impose the following normalization
\begin{equation*}
{F}_{N}(\zeta\cdot\eta):=\tilde{C}(N)\tilde{F}_{N}(\zeta\cdot\eta),
\end{equation*}
with $\widetilde{C}(N)>0$, chosen such that
\begin{equation} \label{low-bound-F}
F_N(1) = 1,
\end{equation} 
\noindent and 
\begin{equation} \label{bound-der-F}
F_N'(1) \ge  \tilde{c}N^2,
\end{equation}
where $\tilde{c}>0$ is a constant independent of $N$. Indeed, $\tilde{c}$ can be bounded from
below by $1/64$ as follows. Since
\[
F_N \left( t \right)=\tilde{{C}}\left( N \right)\sum\limits_{n=0}^N {\rho 
\left( {\frac{n}{N}} \right)\frac{2n+1}{4\pi }P_{n,3} \left( t \right)}, 
\]
and $P_{n,3}(1)=1$, $\forall n\ge0$, the normalization $F_N \left( 1 \right)=1$, gives
\[
\tilde{{C}}\left( N \right)=\frac{1}{\sum\limits_{n=0}^N {\rho \left( 
{\frac{n}{N}} \right)\frac{2n+1}{4\pi }} }.
\]
The derivative formula (see e.g. \cite{atkinson_spherical})
\[
P'_{n,d}(t)=\frac{n(n+d-2)}{d-1} P_{n-1,d+2}(t), \qquad n \ge 1, d \ge 2,
\]
implies
\[
 {F}'_N \left( t \right)=\tilde{{C}}\left( N \right)\sum\limits_{n=1}^N {\rho \left( {\frac{n}{N}} 
\right)\frac{2n+1}{4\pi }\frac{n\left( {n+1} \right)}{2}P_{n-1,5} \left( t \right)}. 
\]
Hence, by the properties of $\rho$ (see (\ref{def-rho}))
\[
\begin{split}
 F_N '\left( 1 \right)&=\frac{\sum\limits_{n=1}^N {\rho \left( {\frac{n}{N}} 
\right)\frac{2n+1}{4\pi }\frac{n\left( {n+1} \right)}{2}} 
}{\sum\limits_{n=0}^N {\rho \left( {\frac{n}{N}} \right)\frac{2n+1}{4\pi }} 
}\, \\ 
& \ge \frac{\sum\limits_{n=1}^{N/2} {n\left( {n+1} \right)\left( {2n+1} 
\right)} }{2\sum\limits_{n=0}^N {\left( {2n+1} \right)} } \\ 
& =\frac{\frac{1}{2}\frac{N}{2}\left( {\frac{N}{2}+1} \right)\left( 
{\frac{1}{4}N^2+\frac{3}{2}N+2} \right)}{2N^2+4N+2}\ge \frac{N^2}{64}. \\ 
 \end{split}
\]

Our construction requires the right form of differentiation. To this end we employ 
the Lie-Algebra structure on the sphere (see Section 4.2.2 in \cite{atkinson_spherical} for more details).
For any $\xi_0 \in \mathbb{S}^2$ , let $D_{\xi_0 ,1} ,D_{\xi_0 ,2} $, be 
the two Lie Algebra matrices associated with the directions of the vectors spanning the tangent plane 
at $\xi_0 \in \mathbb{S}^2$. The two tangents and hence the matrices, can be determined uniquely (and continuously) to form a right-hand system with $\xi_0$. These matrices generate parametric families of rotation at angles $t$ in the
corresponding directions by the rotation matrices 
\begin{equation*}
D_{\xi_0 ,1} \left( t \right):=e^{-tD_{\xi_0 ,1} },D_{\xi_0 ,2} \left( t 
\right):=e^{-tD_{\xi_0 ,2} },
\end{equation*}
\noindent where for any matrix $B$, $e^B:=\sum_0^\infty {\frac{B^k}{k!}}$. We may define the rotational 
derivatives (if exist) of a function $F:\mathbb{S}^2 \rightarrow \mathbb{R}$, at a point $\xi \in \mathbb{S}^2$, by
\[
D_{\xi_0 ,r}F(\xi):= \lim\limits_{t\to 
0} \frac{F\left( {D_{\xi_0 ,r} \left( t \right)\xi}\right)-F \left( {\xi} \right)}{t}, \qquad r=1,2.
\]
Thus, for any point $\xi_1 \in \mathbb{S}^2$, we define the rotational derivatives associated with $\xi_0$, of the 
function $F_N(\xi \cdot \xi_1)$, localized at $\xi_1$, by

\begin{align*}
D_{\xi_0 ,1}& F_N \left( {\xi,\xi_1 } \right):=\lim\limits_{t\to 
0} \frac{F_N \left( {D_{\xi_0 ,1} \left( t \right)\xi \cdot \xi_1 } 
\right)-F_N \left( {\xi \cdot \xi_1 } \right)}{t},\\
D_{\xi_0 ,2}& F_N \left( {\xi,\xi_1 } \right):=\lim\limits_{t\to 
0} \frac{F_N \left( {D_{\xi_0 ,2} \left( t \right)\xi \cdot \xi_1 } 
\right)-F_N \left( {\xi \cdot \xi_1 } \right)}{t}.
\end{align*}
Denoting briefly $\mathcal{P}$ as the orthogonal projector onto $V_N$, we know by Lemma 4.7 of
\cite{atkinson_spherical} that for any polynomial $Q\in V_N$,
\[
D_{\xi_0 ,r} Q = D_{\xi_0 ,r} \mathcal{P} Q = \mathcal{P}D_{\xi_0 ,r} Q, \qquad r=1,2,
\]
\noindent which implies that $D_{\xi_0 ,r} F_N \left( {\xi\cdot \xi_1 } \right)\in V_N$, $r=1,2$, i.e. are spherical harmonics. This is crucial for the construction of the interpolating polynomial (\ref{eq:q_form}).  

First, we investigate the properties of the spherical harmonic $G(\xi,\xi_0):= \xi \cdot \xi_0$,
for a fixed $\xi_0\in\mathbb{S}^{d-1}$

\begin{lemma} \label{dot-diff-estimate} For any $\xi_0, \eta ,\eta_1 ,\eta_2 \in \mathbb{S}^{d-1}$
\[
\left| {G(\eta_1,\xi_0) -G(\eta_2,\xi_0) } \right|\le d\left( 
{\eta_1 ,\eta_2 } \right)\left[ {d\left( {\eta ,\xi_0 } \right)+ \max_{j=1,2} d(\eta,\eta_j)} \right].
\]

\begin{proof}
Denote $d_1 :=d\left( {\eta_1 ,\xi_0 } \right),d_2 :=d\left( {\eta_2 
,\xi_0 } \right)$ . Then
\[
\begin{split}
 \left| {\eta_1 \cdot \xi_0 -\eta_2 \cdot \xi_0 } \right|&=\left| {\cos 
d_1 -\cos d_2 } \right|=2\left| {\sin \left( {{\left( {d_1 -d_2 } \right)} 
\mathord{\left/ {\vphantom {{\left( {d_1 -d_2 } \right)} 2}} \right. 
\kern-\nulldelimiterspace} 2} \right)} \right|\left| {\sin \left( {{\left( 
{d_1 +d_2 } \right)} \mathord{\left/ {\vphantom {{\left( {d_1 +d_2 } 
\right)} 2}} \right. \kern-\nulldelimiterspace} 2} \right)} \right| \\ 
& \le 1/2\left| {d_1 -d_2 } \right|\left| {d_1 +d_2 } \right|. \\ 
\end{split}
\]
Hence
\[
\begin{split}
 \left| {\eta_1 \cdot \xi_0 -\eta_2 \cdot \xi_0 } \right|&\le d\left( 
{\eta_1 ,\eta_2 } \right)\max \left\{ {d\left( {\eta_1 ,\xi_0 } 
\right),d\left( {\eta_2 ,\xi_0 } \right)} \right\} \\ 
&\le d\left( {\eta_1 ,\eta_2 } \right)\left( {d\left( {\eta ,\xi_0 } 
\right)+\max_{j=1,2} d(\eta,\eta_j)} \right). \\ 
\end{split}
\]

\end{proof}
\end{lemma}

\noindent Let $\xi_0,\xi_1,\eta \in \mathbb{S}^2$, $r=1,2$ and $0<t \le \pi$. If $D_{\xi_1,r}(t)\eta=\eta$,
then obviously $D_{\xi_1,r}G(\eta,\xi_0)=0$. Else, observe that for any rotation matrix $A$, at an angle $t$, 
applied to $\eta$, we have $d(A\eta,\eta) \le t$. Applying this observation and Lemma \ref{dot-diff-estimate},
give 
\begin{equation} \label{lim-dist}
\begin{split}
|D_{\xi_1,r}G(\eta,\xi_0)|&= \lim\limits_{t\to 0} \frac{\left| {D_{\xi_1 ,r} \left( t \right)\eta 
\cdot \xi_0 -\eta \cdot \xi_0 } \right|}{t} \\
&\le \lim\limits_{t\to 0} 
\frac{d\left( {D_{\xi_1 ,r} \left( t \right)\eta ,\eta } \right)\left( 
{d\left( {\eta ,\xi_0 } \right)} + d\left( {D_{\xi_1 ,r} \left( t \right)\eta ,\eta } \right) \right)}{d\left( {D_{\xi_1 ,r} \left( t \right)\eta ,\eta } \right)} \\ 
& \le d\left( {\eta ,\xi_0 } \right). \\ 
 \end{split}
\end{equation}
Next, we have the Lipschitz-type estimate
\begin{equation} \label{G-1}
\begin{split}
 \left| {D_{\xi_1 ,r } G\left( {\eta_1 ,\xi_0 } \right)-D_{\xi_1 ,r 
} G\left( {\eta_2 ,\xi_0 } \right)} \right|&=\lim\limits_{t\to 0} 
\frac{\left| {\left( {D_{\xi_1 ,r } \left( t \right)-I} \right)\left( 
{\eta_1 -\eta_2 } \right)\cdot \xi_0 } \right|}{t} \\ 
& \le \lim\limits_{t\to 0} 
\frac{||D_{\xi_1 ,r }(t)-I|| |\eta_1 -\eta_2| |\xi_0|}{t} \\ 
& \le \left| {\eta_1 -\eta_2 } \right| \\
& \le d\left( {\eta_1 ,\eta_2 } 
\right). \\ 
 \end{split}
\end{equation}
This gives for any $\xi_0,\xi_1,\xi_2,\eta \in \mathbb{S}^2$
\begin{equation} \label{G-2}
|D_{\xi_1,r_1}D_{\xi_2,r_2}G(\eta,\xi_0)| \le 1.
\end{equation}
We now recall the following estimate for every $k \ge 1$, $\ell \geq 0$ and $\zeta,\eta\in \mathbb{S}^{d-1}$
 \cite{petrushev_decomposition},
\begin{equation}
\Bigg\vert F_N^{(\ell)}(\zeta \cdot \eta)\Bigg\vert\leq \frac{c_{k,\ell}N^{2\ell}}{(1+Nd(\zeta,\eta))^k}, \label{eq:bound}
\end{equation}
where $c_{k,\ell}$ is a positive constant depending only on $k $, $\ell$. This already gives the good localization 
of $F_N(\xi \cdot \xi_0)$ at $\xi_0 \in \mathbb{S}^2$, for any $k\ge 1$ 
\begin{equation} \label{eq:bound-FN}
\left| {F_N \left( {\xi \cdot \xi_0 } \right)} \right|\le \frac{c_k 
}{\left( {1+Nd\left( {\xi ,\xi_0 } \right)} \right)^k}.
\end{equation}
Let us proceed with localization of derivatives. For any $\xi_0 ,\xi_1 \in \mathbb{S}^2$ and $r=1,2$ we have 
the following chain rule
\[
\begin{split}
 D_{\xi_1 ,r} F_N \left( {\xi ,\xi_0 } \right)&=\lim\limits_{t\to 0} 
\frac{F_N \left( {D_{\xi_1 ,r} \left( t \right)\xi \cdot \xi_0 } 
\right)-F_N \left( {\xi \cdot \xi_0 } \right)}{t} \\ 
 &=\lim\limits_{t\to 0} \frac{\left( {F_N \left( {D_{\xi_1 ,r} \left( t 
\right)\xi \cdot \xi_0 } \right)-F_N \left( {\xi\cdot \xi_0 } \right)} 
\right)}{D_{\xi_1 ,r} \left( t \right)\xi \cdot \xi_0 -\xi \cdot \xi_0 
}\frac{D_{\xi_1 ,r} \left( t \right)\xi \cdot \xi_0 -\xi \cdot \xi_0 }{t} \\ 
& ={F}'_N \left( {\xi \cdot \xi_0 } \right)D_{\xi_1,r}G(\xi,\xi_0). \\ 
\end{split}
\]
We note that the above representation of the derivative also shows that it is 
a spherical polynomial of degree $\le N$. Furthermore, in the special case where $\xi =\xi_0 =\xi_1 $, we get
\begin{equation} \label{zero-der-1-1}
\begin{split}
 D_{\xi_0 ,r} F_N \left( {\xi_0 ,\xi_0 } \right)&={F}'_N \left( 1 
\right)\lim\limits_{t\to 0} \frac{D_{\xi_0 ,r} \left( t \right)\xi_0 
\cdot \xi_0 -1}{t} \\ 
&={F}'_N \left( 1 \right)\lim\limits_{t\to 0} \frac{\cos t-1}{t}=0. \\ 
\end{split}
\end{equation}
We require the following result that generalizes a lemma from \cite{petrushev_decomposition}

\begin{lemma} \label{lemma:lip-1} Let $\xi_0,\eta ,\eta_1 ,\eta_2 \in \mathbb{S}^2$ with $d\left( {\eta_j 
,\eta } \right)\le N^{-1}$, $j=1,2$. Then, for any $k\ge1,\ell\ge0$,
\begin{equation} \label{eq:lip-1}
\left| {F_N^{(\ell)} \left( {\eta_1 \cdot \xi_0 } \right)-F_N^{(\ell)} \left( {\eta_2 \cdot \xi_0 } 
\right)} \right|\le \frac{c_{k,\ell} d\left( {\eta_1 ,\eta_2 } 
\right)N^{2\ell+1}}{\left( {1+Nd\left( {\eta ,\xi_0 } \right)} \right)^k},
\end{equation}

\begin{proof} First observe that by the triangle inequality for any $\tilde{\eta}$ such that
$d(\tilde{\eta},\eta) \le N^{-1}$
\[
\begin{split}
 N d \left( {\eta ,\xi_0 } \right)&\le N\left( {d \left( {\eta 
,\tilde{{\eta }}} \right)+ d \left( {\tilde{{\eta }},\xi_0 } \right)} 
\right) \\ 
& \le N\left( {N^{-1}+d \left( {\tilde{{\eta }},\xi_0 } \right)} 
\right) \\ 
& \le 1 +N d \left( {\tilde{{\eta }},\xi_0 } \right). \\ 
\end{split}
\]
Applying this, (\ref{eq:bound}) and Lemma \ref{dot-diff-estimate} yields 
\[
\begin{split}
 \left| {F_N^{(\ell)} \left( {\eta_1 \cdot \xi_0 } \right)-F_N^{(\ell)} \left( {\eta_2 
\cdot \xi_0 } \right)} \right|&\le \max_{d(\tilde{\eta},\eta) \le N^{-1}} \left| {F^{(\ell+1)}_N \left( 
{\tilde{{\eta }}\cdot \xi_0 } \right)} \right|\left| {\eta_1 \cdot \xi_0 
-\eta_2 \cdot \xi_0 } \right| \\ 
&\le \frac{c_{k+1,\ell+1} N^{2(\ell+1)}}{\left( {1+Nd\left( {\tilde{\eta} ,\xi_0 } \right)} 
\right)^{k+1}}d\left( {\eta_1 ,\eta_2 } \right)\left[ {d\left( {\eta ,\xi 
_0 } \right)+N^{-1}} \right] \\ 
&\le \frac{cN^{2\ell+1}d\left( {\eta_1 ,\eta_2 } \right)}{\left( {1+Nd\left( {\eta 
,\xi_0 } \right)} \right)^k}+\frac{cN^{2\ell+1}d\left( {\eta_1 ,\eta_2 } 
\right)}{\left( {1+Nd\left( {\eta ,\xi_0 } \right)} \right)^{k+1}} \\ 
 &\le \frac{cN^{2\ell+1}d\left( {\eta_1 ,\eta_2 } \right)}{\left( {1+Nd\left( {\eta 
,\xi_0 } \right)} \right)^k}. \\ 
\end{split}
\]

\end{proof}
\end{lemma}

\noindent As a conclusion from Lemma \ref{lemma:lip-1}, we obtain the localization of the derivatives, i.e. for any $\xi_0,\xi_1 \in \mathbb{S}^2$ and $r=1,2$ 
\begin{equation} \label{eq:estimate-diff-1}
\begin{split}
 \left| {D_{\xi_1 ,r} F_N \left( {\xi ,\xi_0 } \right)} \right|&=\lim\limits_{t\to 0} \frac{\left| {F_N \left( {D_{\xi_1 ,r} \left( t \right)\xi \cdot \xi_0 } \right)-F_N \left( {\xi \cdot \xi_0 } \right)} \right|}{t} \\ 
&\le \lim\limits_{t\to 0} \frac{\left| {F_N \left( {D_{\xi_1 ,r} \left( 
t \right)\xi \cdot \xi_0 } \right)-F_N \left( {\xi \cdot \xi_0 } \right)} 
\right|}{d\left( {D_{\xi_1 ,r} \left( t \right)\xi ,\xi } \right)} \\ 
& \le \frac{c_k N}{\left( {1+Nd\left( {\xi ,\xi_0 } \right)} \right)^k}. \\ 
\end{split}
\end{equation}

\noindent Next, we analyze second order derivatives. By the rotation invariance of functions of the type $F_N(\xi \cdot \xi_0)$, we may compute certain values of partial derivatives at the point $\xi_0 =\left( {-1,0,0} \right)$.  The rotations at the angle $t$ associated with the partial derivatives at $\xi_0$ are
\[
D_{\xi_0 ,1} \left( t \right)=\left( {{\begin{array}{*{20}c}
 {\cos t} \hfill & 0 \hfill & {\sin t} \hfill \\
 0 \hfill & 1 \hfill & 0 \hfill \\
 {-\sin t} \hfill & 0 \hfill & {\cos t} \hfill \\
\end{array} }} \right),
\quad
D_{\xi_0 ,2} \left( t \right)=\left( {{\begin{array}{*{20}c}
 {\cos t} \hfill & {\sin t} \hfill & 0 \hfill \\
 {-\sin t} \hfill & {\cos t} \hfill & 0 \hfill \\
 0 \hfill & 0 \hfill & 1 \hfill \\
\end{array} }} \right).
\]
Let $G(\eta)=G(\eta_1,\eta_2,\eta_3):\mathbb{R}^3\to \mathbb{R}$, be any differentiable function. Following Section 1.8 in \cite{daixu},
we compute for $\xi_0 =\left( {-1,0,0} \right)$ and $\eta \in \mathbb{S}^2$
\begin{equation} \label{eq:RepD1}
\begin{split}
 D_{\xi_{0} ,1} G\left( \eta \right)&=\lim\limits_{t\to 0} \frac{G\left( 
{D_{\xi_{0} ,1} \left( t \right)\eta } \right)-G\left( \eta \right)}{t} \\ 
 &=\eta_{3} \partial_{1} G\left( \eta \right)-\eta_{1} \partial_{3} 
G\left( \eta \right). \\ 
\end{split}
\end{equation}
Similarly 
\begin{equation} \label{eq:RepD2}
\begin{split}
 D_{\xi_{0} ,2} G\left( \eta \right)&=\lim\limits_{t\to 0} \frac{G\left( 
{D_{\xi_{0} ,2} \left( t \right)\eta } \right)-G\left( \eta \right)}{t} \\ 
&=\eta_{2} \partial_{1} G\left( \eta \right)-\eta_{1} \partial_{3} 
G\left( \eta \right). \\ 
\end{split}
\end{equation}
In the special case $G(\eta):=F_{N}(\eta \cdot \xi_{0})$ , with $\xi_{0} =( -1,0,0)$, we obtain by (\ref{eq:RepD1})
\[
\begin{split}
 D_{\xi_{0} ,1} F_{N} \left( {\eta ,\xi_{0} } \right)&=\eta_{3} \partial_{1} F_{N} \left( {-\eta_{1} } \right)-\eta_{1} 
\partial_{3} F_{N} \left( {-\eta_{1} } \right) \\ 
&=-\eta_{3} {F}'_{N} \left( {-\eta_{1} } \right)= -\eta_{3} {F}'_{N} 
\left( {\eta \cdot \xi_{0} } \right). \\ 
\end{split}
\]
Applying (\ref{eq:RepD2}) with similar computation gives
\[
D_{\xi_{0} ,2} F_{N} \left( {\eta ,\xi_{0} } \right)=-\eta_{2} {F}'_{N} 
\left( {\eta \cdot \xi_{0} } \right).
\]
This correlates with what we already observed (see (\ref{zero-der-1-1})), namely, that for any $\xi_0 \in \mathbb{S}^2$
\begin{equation} \label{zero-part-der}
D_{\xi_0 ,1} F_N \left( {\xi_0, \xi_0 } \right) = D_{\xi_0 ,2} F_N \left( {\xi_0, \xi_0 } \right)=0.
\end{equation}

\noindent Using (\ref{eq:RepD1}) and (\ref{eq:RepD2}), we may compute mixed partial derivatives at $\xi_0=(-1,0,0)$, 
\[
 D_{\xi_0 ,2} D_{\xi_0 ,1} F_N \left( {\eta, \xi_0 } \right) = \eta_2 \eta_3 {F}''_N \left( {\eta \cdot \xi_0 } \right).
\]
This implies that for $\xi_0=(-1,0,0)$, 
\begin{equation} \label{mix-part-zero}
D_{\xi_0 ,2} D_{\xi_0 ,1} F_N ( {\xi_0, \xi_0 } ) = D_{\xi_0 ,1} D_{\xi_0 ,2} F_N ( {\xi_0, \xi_0 } )=0,
\end{equation}
but obviously, by the rotation invariance, (\ref{mix-part-zero}) holds for any point $\xi_0 \in \mathbb{S}^2$. We also get for $\xi_0=(-1,0,0)$ using (\ref{eq:RepD1}) and (\ref{eq:RepD2}) 
\[
 D_{\xi_0 ,2} D_{\xi_0 ,2} F_N \left( {\eta, \xi_0 } \right) =\eta_1 {F}'_N \left( {\eta \cdot \xi_0 } \right)+\eta_2^2 {F}''_N \left( 
{\eta \cdot \xi_0 } \right). 
\]
With similar computations for $D_{\xi_0 ,1} D_{\xi_0 ,1} F_N$, and the rotation invariance, we have for any $\xi_0 \in \mathbb{S}^2$
\begin{equation} \label{2-ord-der-diag}
D_{\xi_0 ,1} D_{\xi_0 ,1} F_N \left( {\xi_0, \xi_0 } 
\right)= D_{\xi_0 ,2} D_{\xi_0 ,2} F_N \left( {\xi_0, \xi_0 } 
\right) = -F'_N(1).
\end{equation}
Proceeding to the next higher order Lipschitz estimate for $\eta,\eta_1,\eta_2\in \mathbb{S}^2$,
satisfying $d(\eta_1,\eta),d(\eta_2,\eta) \le N^{-1}$, we have
\[
\begin{array}{l}
 D_{\xi_1 ,r} F_N \left( {\eta_1 ,\xi_0 } \right)-D_{\xi_1 ,r} F_N 
\left( {\eta_2 ,\xi_0 } \right) \\
\qquad ={F}'_N \left( {\eta_1 \cdot \xi_0 } 
\right)D_{\xi_1 ,r}G(\eta_1,\xi_0)-{F}'_N \left( {\eta_2 \cdot \xi_0 
} \right)D_{\xi_1 ,r}G(\eta_2,\xi_0) \\ 
\qquad = \left( { {F}'_N(\eta_1 \cdot \xi_0) - {F}'_N (\eta_2 \cdot \xi_0)} \right)
D_{\xi_1 ,r}G(\eta_1,\xi_0) +{F}'_N \left( {\eta_2 \cdot \xi_0 } \right)(D_{\xi_1 ,r}G(\eta_1,\xi_0)-D_{\xi_1 ,r}G(\eta_2,\xi_0)).\\ 
\end{array}
\]
Consequently, using (\ref{lim-dist}),(\ref{G-1}), (\ref{eq:bound}) and (\ref{eq:lip-1}) for $\ell=1$ yields
\begin{equation} \label{eq:lip-2}
\begin{array}{l}
 \left| {D_{\xi_1 ,r} F_N \left( {\eta_1 ,\xi_0 } \right)-D_{\xi_1 ,r 
} F_N \left( {\eta_2 ,\xi_0 } \right)} \right| \\
\quad \le \left| {{F}'_N \left( 
{\eta_1 \cdot \xi_0 } \right)-{F}'_N \left( {\eta_2 \cdot \xi_0 } 
\right)} \right|d\left( {\eta_1 ,\xi_0 } \right)+\left| {{F}'_N \left( 
{\eta_2 \cdot \xi_0 } \right)} \right|d\left( {\eta_1 ,\eta_2 } \right) \\ 
\quad \le \frac{c_{k+1} N^3}{\left( {1+Nd\left( {\eta \cdot \xi_0 } \right)} 
\right)^{k+1}}d\left( {\eta_1 ,\eta_2} \right) \left( {d(\eta ,\xi_0)+N^{-1} } \right)+\frac{c_k N^2d\left( {\eta 
_1 ,\eta_2 } \right)}{\left( {1+Nd\left( {\eta ,\xi_0 } \right)} 
\right)^k} \\ 
\quad \le \frac{c_k N^2d\left( {\eta_1 ,\eta_2 } \right)}{\left( {1+Nd\left( 
{\eta ,\xi_0 } \right)} \right)^k}. \\ 
 \end{array}
\end{equation}
This implies for any $\xi_0 ,\xi_1 ,\xi_2\in \mathbb{S}^2$, $r_1 ,r_2 =1,2$,
\begin{equation} \label{eq:estimate-diff-2}
\left| {D_{\xi_2 ,r_2 } D_{\xi_1 ,r_1 } F_N \left( {\xi ,\xi_0 } \right)} 
\right|\le \frac{c_k N^2}{\left( {1+Nd\left( {\xi ,\xi_0 } \right)} 
\right)^k}.
\end{equation}
Similar calculations give
\begin{equation} \label{eq:lip-3}
\left| {D_{\xi_1 ,r_1 } D_{\xi_2 ,r_2 } F_N \left( {\eta_1 ,\xi_0 } 
\right)-D_{\xi_1 ,r_1 } D_{\xi_2 ,r_2 } F_N \left( {\eta_2 ,\xi_0 } \right)} \right|\le 
\frac{c_k N^3d\left( {\eta_1 ,\eta_2 } \right)}{\left( {1+Nd\left( {\eta 
,\xi_0 } \right)} \right)^k},
\end{equation}
which in turn yields for any $\xi_0 ,\xi_1 ,\xi_2,\xi_3 \in \mathbb{S}^2$ , $r_1 ,r_2,r_3=1,2$,
\begin{equation} \label{eq:third-der}
\left| {D_{\xi_1 ,r_1 }D_{\xi_2 ,r_2 } D_{\xi_3 ,r_3 } F_N \left( {\xi \cdot \xi_0 } 
\right)} \right|\le \frac{c_k N^3}{\left( {1+Nd \left( {\xi,\xi_0 
} \right)} \right)^k}.
\end{equation}

\section{The construction of the interpolating polynomial on $\mathbb{S}^2$} \label{sec:construct}
According to Theorem \ref{th:duality}, a sufficient condition for the recovery 
of $f$ from its `orthogonal projection' onto $V_N(\mathbb{S}^2)$ is the existence of $q\in V_N$, satisfying 
\begin{align}
q(\mathbf{\xi}_m)&=u_m, \quad \forall \xi_m\in\Xi, \label{eq:con1}\\
\vert q(\mathbf{\xi})\vert &<1, \quad \forall \xi\notin\Xi ,\label{eq:con2}
\end{align}
for any {signed} sequence $\{u_m\}$ with $|u_m|=1$. Following the construction of
\cite{candes2013towards} for $d=2$, we propose that the appropriate form for $d=3$ is
\begin{equation} 
q\left( \xi \right):=\sum\limits_{\xi_m\in\Xi} {\alpha_m F_N \left( {\xi \cdot \xi_m } 
\right)+\beta_m D_{\xi_m ,1} F_N \left( {\xi, \xi_m } \right)+\gamma 
_m D_{\xi_m ,2} F_N \left( {\xi, \xi_m } \right)} ,
 \label{eq:q_form}
\end{equation}
where $\{\alpha_m\}$,$\{\beta_m\}$, and $\{\gamma_m\}$ are sequences of real coefficients, to be selected
later. We point out that, as explained in Section \ref{sec:loc}, the partial derivatives in (\ref{eq:q_form}) are spherical harmonics polynomials of degree $\leq N$, and thus $q\in V_N(\mathbb{S}^2)$. 

Thus, this section is devoted to the proof of the following proposition:
\begin{proposition}
If $\Xi \subset \mathbb{S}^2$ satisfies the separation condition of Definition \ref{def:separation}, then there exist coefficients
$\{\alpha_m\}$,$\{\beta_m\}$, and $\{\gamma_m\}$ such that $q$ of the form (\ref{eq:q_form}) obeys (\ref{eq:con1}) and (\ref{eq:con2}). 
\label{prop:1}
\end{proposition}

According to Theorem \ref{th:duality}, Proposition \ref{prop:1} immediately implies Theorem \ref{th:main}. 
The proof of Proposition \ref{prop:1} follows the outline of \cite{candes2013towards} 
and is given by a series of lemmas, as follows:

\begin{lemma}
If  the separation condition of Definition \ref{def:separation} holds, then 
for any sequence $\{ u_m \}$, with $u_m= \{-1,1 \}$, there exist coefficients $\{\alpha_m\}$,$\{\beta_m\}$, and $\{\gamma_m\}$, such that 
\begin{align}
q({\xi}_m)&=u_m, \label{eq:con3}\\
{D_{\xi_m,1}q(\xi_m)}& = {D_{\xi_m,2}q(\xi_m)}=0,   \label{eq:con4}
\end{align}
for all $\xi_m\in\Xi$. Additionally, for any $k \ge 3$, there exists a constant $c_k$, such that
\begin{align}
\|\alpha\|_\infty&\leq 1+\frac{c_k}{\nu^{k-1}}, \label{bound-alpha} \\
\|\beta\|_\infty&\leq \frac{c_k}{N\nu^{k-1}}, \label{bound-beta} \\
\|\gamma\|_\infty&\leq \frac{c_k}{N\nu^{k-1}}, \label{bound-gamma}
\end{align}
with $\nu >0$, the constant from the separation condition. Moreover, if $u_1=1$, then 
\begin{equation}
\alpha_1\geq 1-\frac{c_k}{\nu^{k-1}}. \label{eq:alpha0}
\end{equation}
\label{lemma:1}
\end{lemma}

\begin{lemma}
If  the separation condition in Definition \ref{def:separation} holds, then  the polynomial (\ref{eq:q_form}) 
as constructed in Lemma \ref{lemma:1} satisfies $\vert q(\xi)\vert <1$ for any $\xi\in \mathbb{S}^2$, obeying  
\begin{equation*}
d\left(\xi,\xi_m\right)\leq\frac{\sigma}{N}, 
\end{equation*}
for some $\xi_m\in\Xi$ and sufficiently small $\sigma>0$.
\label{lemma:2}
\end{lemma}

\begin{lemma}
If  the separation condition in Definition \ref{def:separation} holds, then  the polynomial  (\ref{eq:q_form}) 
as constructed in Lemma \ref{lemma:1} satisfies $\vert q(\xi)\vert <1$ for any $\xi\in \mathbb{S}^2$, obeying 
\begin{equation*}
d\left(\xi,\xi_m\right)\ge \frac{\sigma}{N}, \quad  \forall \xi_m\in\Xi,
\end{equation*}
where $\sigma$ is the constant of Lemma \ref{lemma:2}.
\label{lemma:3}
\end{lemma}

\subsection{Proof of Lemma \ref{lemma:1}}

\noindent The gradient of any $q$ of the form (\ref{eq:q_form}), at a point $\xi_k\in\Xi$, is given by
\[
\begin{split}
 D_{\xi_k ,r} q\left( {\xi_k } \right)&=\sum\limits_{\xi_m \in \Xi } 
{\alpha_m D_{\xi_k ,r} F_N \left( {\xi_k, \xi_m } \right)+
\beta_m D_{\xi_k ,r} D_{\xi_m ,1} F_N\left( {\xi_k, \xi_m } \right)} \\ 
& \qquad \qquad +\gamma_m D_{\xi_k ,r} D_{\xi_m ,2} F_N 
\left( {\xi_k, \xi_m } \right), \qquad \qquad r=1,2. \\ 
\end{split}
\]

\noindent Conditions (\ref{eq:con3}) and (\ref{eq:con4}) may be written in matrix notation as
\begin{align}
 \begin{bmatrix}
F_0 & \tilde{F}_1^1 & \tilde{F}_1^2 \\ F_1^1 & F_2^{1,1} &  F_2^{1,2}\\ F_1^2 &  F_2^{2,1} &  F_2^{2,2}
\end{bmatrix}
\begin{bmatrix}
\alpha \\  \beta\\ \gamma
\end{bmatrix}
=\begin{bmatrix}
u \\  0  \\ 0
\end{bmatrix}, \label{eq:big_matrix}
\end{align}
where
\begin{align*}
&F_0:=\left\{ {F_N \left( {\xi_k \cdot \xi_m } \right)} \right\}_{k,m}, \\
&F_1^r:=\left\{ {D_{\xi_k ,r} F_N \left( {\xi_k, \xi_m } \right)} \right\}_{k,m}, \quad r=1,2,\\
&\tilde{F}_1^r:=\left\{ {D_{\xi_m ,r} F_N \left( {\xi_k, \xi_m } \right)} \right\}_{k,m}, \quad r=1,2,\\
&F_2^{r_1 ,r_2 }:=\left\{ {D_{\xi_k ,r_1} D_{\xi_m ,r_2} F_N 
\left( {\xi_k, \xi_m } \right)} \right\}_{k,m}, \quad r_1,r_2=1,2,
\end{align*}
and $u=\{u_m\}_m, \alpha=\{\alpha_m\}_m, \beta=\{\beta_m\}_m,  \gamma=\{\gamma_m\}_m$. 
 For convenience, we occasionally write (\ref{eq:big_matrix}) as 
 \begin{align*}
 \mathcal{F}=
 \begin{bmatrix}
{F}_0 & \tilde{\mathcal{F}}_1  \\\mathcal{F}_1 & \mathcal{F}_2 
\end{bmatrix}.
\end{align*}
Our goal is to show that $\mathcal{F}$ is invertible and to estimate 
the coefficients $\alpha,\beta,\gamma$. To this purpose, we require the following 

\begin{lemma} \label{lemma-sum-estimate}

Let $\xi_0 \in \Xi$, where $\Xi$ satisfies the separation condition and let $\xi \in \mathbb{S}^2$, 
such that $d(\xi,\xi_0)\leq \Delta/2$. Then, for any $k \ge 3$ there exists $c_k>0$, such that for 
any $\tilde{\xi}_1,\tilde{\xi}_2,\tilde{\xi}_3\in \mathbb{S}^2$ and $r_1,r_2,r_3=1,2$, 
\begin{align}
 \sum_{\xi_m \in \Xi \backslash \xi_0 } {\left| {F_N \left( {\xi
\cdot\xi_m } \right)} \right|} &\le \frac{c_k }{\nu^{k-1}},\label{point-kernel-estimate_0} \\ 
 \sum_{\xi_m \in \Xi \backslash \xi_0 } {\left| {D_{\tilde{\xi}_1 ,r_1} F_N 
\left( {\xi,\xi_m } \right)} \right|},  \sum_{\xi_m \in \Xi \backslash \xi_0 } {\left| {D_{\xi_m ,r_1} F_N 
\left( {\xi,\xi_m } \right)} \right|} &\le \frac{c_k N}{\nu^{k-1}}, \label{point-kernel-estimate_1} \\ 
 \sum_{\xi_m \in \Xi \backslash \xi_0 } {\left| {D_{\tilde{\xi}_1 ,r_1 } 
D_{\tilde{\xi}_2 ,r_2 } F_N \left( {\xi,\xi_m } \right)} \right|} &\le \frac{c_k 
N^2}{\nu^{k-1}},  \label{point-kernel-estimate_2} \\
 \sum_{\xi_m \in \Xi \backslash \xi_0 } {\left| {D_{\tilde{\xi}_1 ,r_1 } 
D_{\tilde{\xi}_2 ,r_2 } D_{\tilde{\xi}_3 ,r_3 }F_N \left( {\xi,\xi_m } \right)} \right|} &\le \frac{c_k 
N^3}{\nu^{k-1}}.  \label{point-kernel-estimate_3}
\end{align}
\begin{proof}

Fix $\xi_0 \in\Xi$. Let $\Omega_m $ be the `ring' about $\xi_0 $ such 
that
\begin{equation*}
\Omega_m :=\left\{ {\xi \in {\mathbb{S}}^2\mbox{\thinspace :\thinspace 
}\frac{\nu m}{N} < d\left( {\xi, \xi_0 } 
\right) \le  \frac{\nu \left( {m+1} \right)}{N}} \right\} , 0\leq m\leq 
\left\lfloor {\frac{\pi N}{\nu }-1} \right\rfloor .
\end{equation*}
The surface area of the ring is given by \cite{atkinson_spherical}
\begin{equation*}
\left| {\Omega_m } \right|=2\pi \left( {\cos \left( {\frac{\nu }{N}m} 
\right)-\cos \left( {\frac{\nu }{N}\left( {m+1} \right)} \right)} \right).
\end{equation*}
By assumption, the set $\Xi$ satisfies the separation condition in Definition \ref{def:separation}. Hence, 
the points are the center of pairwise disjoint caps of area $2\pi \left( 
{1-\cos \frac{\nu }{2N}} \right)$. 
Observe that the cap of any $\xi_k \in \Omega_m $ is contained in the ring
\begin{equation*}
\widetilde{{\Omega }}_m :=\left\{ {\xi \in \mathbb{S}^2\mbox{\thinspace 
:\thinspace }\max \left\{ {\frac{\nu \left( {m-1/2} \right)}{N},0} \right\} < d\left( 
{\xi , \xi_0 } \right) \le \min \left\{ {\frac{\nu \left( {m+3/2} \right)}{N},\pi} \right\} } \right\}.
\end{equation*}
Therefore, we can bound the number of points in the ring $\Omega_m $, by
\begin{equation}
\begin{split}
 \# \left\{ {\xi_k \in \Omega_m } \right\}&\leq
  \frac{\left| 
{\tilde{{\Omega }}_m } \right|}{2\pi \left( {1-\cos \frac{\nu }{2N}} 
\right)} 
 =\frac{2\pi \left( {\cos \left( {\frac{\nu }{N}\left( {m-1/2} \right)} 
\right)-\cos \left( {\frac{\nu }{N}\left( {m+3/2} \right)} \right)} 
\right)}{2\pi \left( {1-\cos \frac{\nu }{2N}} \right)} \\
 & =\frac{\sin \left( {\frac{\nu }{2N}\left( {2m+1} \right)} \right)\sin 
\left( {\frac{\nu }{N}} \right)}{\sin^2\left( {\frac{\nu }{4N}} \right)} 
  \leq\frac{\sin \left( {\frac{\nu }{2N}\left( {2m+1} \right)} \right)4\sin 
\left( {\frac{\nu }{4N}} \right)}{\sin 
^2\left( {\frac{\nu }{4N}} \right)} \\
& \leq 4\left| {\frac{\sin \left( {\frac{\nu }{2N}\left( {2m+1} \right)} 
\right)}{\sin \left( {\frac{\nu }{4N}} \right)}} \right| 
 \leq c m, \label{eq:ring_bound}
 \end{split}
\end{equation}

\noindent where the constant does not depend on $N$ or $\nu$. Since $d(\xi,\xi_0)\leq \Delta/2$,
the point $\xi$ is well-separated from the points $\xi_m \in \Xi \backslash \xi_0$. Therefore, 
using (\ref{eq:bound-FN}) and (\ref{eq:ring_bound}) we get
for $k \ge 3$
\begin{equation*}
\begin{split}
\sum\limits_{\xi_j \in \Xi \backslash \xi_0 } {\left| {F_N \left( {\xi 
\cdot \xi_j } \right)} \right|} &\le c_k \sum\limits_{m=1}^\infty {\frac{m}{\left( {1+m\nu } \right)^k}} \\
&\le \frac{c_k }{\nu^{k-1}}\sum\limits_{m=1}^\infty {\frac{1}{m^{k-1}}} \le 
\frac{c_k }{\nu^{k-1}}. \\ 
\end{split}
\end{equation*}

\noindent This proves (\ref{point-kernel-estimate_0}). Using (\ref{eq:estimate-diff-1}), similar 
calculations prove (\ref{point-kernel-estimate_1}) by

\[
\begin{split}
 \sum\limits_{\xi_j \in \Xi \backslash \xi_0 } {\left| {D_{\tilde{\xi}_1 ,r_1 } 
F_N \left( {\xi ,\xi_j } \right)} \right|} &\le c_k N\sum\limits_{m=1}^\infty {\frac{m}{\left( {1+m\nu } \right)^k}} \\ 
&\le \frac{c_k N}{\nu^{k-1}}. \\ 
 \end{split}
\]
The estimates (\ref{point-kernel-estimate_2}) and (\ref{point-kernel-estimate_3}) are proved in a similar manner.

\end{proof} \label{lemma:2.5}
\end{lemma}

\noindent We successively use the fact that a  sufficient condition for the invertibility of a matrix $M$ is 
\begin{equation}
\|I-M\|_\infty<1, \label{eq:inf_norm}
\end{equation}
where $\|M\|_\infty:=max_i\sum_j\vert m_{i,j}\vert$. Furthermore (see e.g \cite{matrix_analysis}, Corollary 5.6.16), 
\begin{equation}
\|M^{-1}\|_\infty\leq\frac{1}{1-\|I-M\|_\infty}. \label{eq:inv_prop}
\end{equation}
 
\noindent The proof of Lemma \ref{lemma:1} also requires the following
\begin{lemma}
If  the separation condition holds, then

\begin{align}
&\left\|I-{F}_0\right\|_\infty\leq\frac{c_k}{\nu^{k-1}} \label{eq:F0}, \\
&\|F_1^r\|_\infty,\|\tilde{F}_1^r\|_\infty\leq N \frac{c_k}{\nu^{k-1}}\label{eq:F1_1}, r=1,2,\\
&\|F_2^{1,2}\|_\infty,\|F_2^{2,1}\|_\infty\leq N^2 \frac{c_k}{\nu^{k-1}} \label{eq:F2_12},\\
&\left\| {-F'_N(1)I-F_2^{r ,r } } \right\|_\infty \leq N^2\frac{c_k }{\nu^{k-1}} \label{eq:F2_22}, \\
&\|(F_2^{r,r})^{-1}\|_\infty \leq  \frac{1}{N^2\left( {\tilde{c}-\frac{c_k }{\nu^{k-1}}} \right)} \quad r=1,2,
 \label{eq:invF2_22}
\end{align}
where the constant $\tilde{c}$ is given by (\ref{bound-der-F}).
\begin{proof}
Observe that by (\ref{low-bound-F}), $F_0(k,k)=F_N(1)=1$. Applying (\ref{point-kernel-estimate_0}) to any row in the matrix $F_0$,
yields (\ref{eq:F0})
\begin{equation*}
\begin{split}
\left\|I-{F}_0\right\|_\infty&=\max_{\xi_j\in\Xi}\sum_{\xi_i\in\Xi,\xi_i\neq \xi_j}\vert F_N(\xi_{j}\cdot\xi_{i})\vert \leq   \frac{c_k}{\nu^{k-1}}.
\end{split}
\end{equation*} 
According to (\ref{zero-part-der}),  the diagonals of $F_1^r$ and $\tilde{F}_1^r$, $r=1,2$ are zero.  Applying (\ref{point-kernel-estimate_1}) gives
\begin{equation*}
\begin{split}
\left\|{F}_1^r\right\|_\infty&=\max_{\xi_j\in\Xi}\sum_{\xi_i\in\Xi,\xi_i\neq \xi_j}\vert {D_{\xi_j ,r}F_N ( {\xi_j, \xi_i })} \vert \leq   \frac{Nc_k}{\nu^{k-1}}.
\end{split}
\end{equation*} 
 In a similar manner, observing from (\ref{mix-part-zero}) that the diagonals of $F_2^{1,2}$ and $F_2^{2,1}$ are zero,
(\ref{point-kernel-estimate_2}) gives (\ref{eq:F2_12}).
Next, we derive from  (\ref{2-ord-der-diag}) and (\ref{point-kernel-estimate_2}) that
\begin{equation*}
\begin{split}
\left\| {-F'_N(1)I-F_2^{r ,r } } \right\|_\infty \leq   \frac{N^2 c_k}{\nu^{k-1}}.
\end{split}
\end{equation*} 
Ultimately, (\ref{eq:inv_prop}), (\ref{eq:F2_22}) and (\ref{bound-der-F}) imply (\ref{eq:invF2_22}).
\end{proof}
\label{lemma:4}
\end{lemma}

\noindent We may now proceed with the proof of Lemma \ref{lemma:1}. To show that $\mathcal{F}$ is invertible for sufficiently large $\nu$, we show that both $\mathcal{F}_2$ and its Schur complement are invertible \cite{zhang2005schur}. 
From (\ref{eq:F2_22}), we know that  $F_2^{2,2}$ is  an invertible matrix for sufficiently large $\nu$. 
So, $\mathcal{F}_2$ is invertible if the Schur complement of $F_2^{2,2}$ in $\mathcal{F}_2$, given by
\begin{equation*}
\mathcal{F}_{s,2}:=(\mathcal{F}_2/F_2^{2,2})=F_2^{1,1}-F_2^{1,2}\left(F_2^{2,2}\right)^{-1}F_2^{2,1},
\end{equation*} 
is invertible as well. Using the estimates of Lemma \ref{lemma:4}, (\ref{bound-der-F}) and assuming
$\nu^{k-1} \ge (1+\tilde{c} c_k)/\tilde{c}^2$, we get
\[
\begin{split}
 \left\| {I-\frac{\mathcal{F}_{s,2}}{-{F}'_N(1)}} 
\right\|_\infty &\le \left\| {I-\frac{F_2^{1,1} }{-{F}'_N \left( 1 \right)}} 
\right\|_\infty +\frac{1}{\left| {{F}'_N \left( 1 \right)} \right|}
\left\| {F_2^{1,2} } \right\|_\infty\left\| {F_2^{2,1} } \right\|_\infty \left\| {\left( {F_2^{2,2} } \right)^{-1}} 
\right\|_\infty \\ 
&\leq\frac{c_k }{\nu^{k-1}}.
\end{split}
\]
This implies that 
\begin{equation}
\|\mathcal{F}_{s,2}^{-1}\|_\infty\leq \frac{1}{F'_N(1)}\frac{1}{1-\frac{c_k}{\nu^{k-1}}}\leq \frac{1}{\tilde{c}N^2}\left(1+\frac{c_k}{\nu^{k-1}-c_k}\right). \label{eq:invF_2s}
\end{equation}
Since $\mathcal{F}_2$ is invertible for sufficiently large $\nu$, 
$\mathcal{F}$ is invertible if the Schur complement $\mathcal{F}_s:=\mathcal{F}/\mathcal{F}_2$  is invertible as well. 
Note that
 \begin{equation*}
 \begin{split}
 (\mathcal{F}/F_2^{2,2})&=\begin{bmatrix} F_0 & \tilde{F}_1^1 \\ F_1^1 &F_2^{1,1} \end{bmatrix}
 -\begin{bmatrix} \tilde{F}_1^2 \\ F_2^{1,2}\end{bmatrix}\left(F_2^{2,2}\right)^{-1}\begin{bmatrix} F_1^2 & F_2^{2,1}\end{bmatrix} \\
&= \begin{bmatrix} F_0 -\tilde{F}_1^2\left(F_2^{2,2}\right)^{-1}F_1^2 & \tilde{\mathcal{F}}_{s,1} \\ \mathcal{F}_{s,1} & \mathcal{F}_{s,2}
\end{bmatrix},
 \end{split}
 \end{equation*}
where 
\begin{align}
\mathcal{F}_{s,1}:=F_1^1-F_2^{1,2}(F_2^{2,2})^{-1}F_1^2, \\
\tilde{\mathcal{F}}_{s,1}:=\tilde{F}_1^1-\tilde{F}_1^{2}(F_2^{2,2})^{-1}F_2^{2,1}.
\end{align}
According to Theorem 1.4 in \cite{zhang2005schur}, 
\begin{equation*}
\mathcal{F}_s=\left(\mathcal{F}/F_2^{2,2}\right)/\left(\mathcal{F}_2/F_2^{2,2}\right),
\end{equation*}
and thus, the Schur complement of $\mathcal{F}_2$ is given by
\begin{equation*}
\mathcal{F}_s=F_0-\tilde{\mathcal{F}}_{s,1}\mathcal{F}_{s,2}^{-1}\mathcal{F}_{s,1}-\tilde{F}_1^2(F_2^{2,2})^{-1}F_1^2.
\end{equation*}
Using Lemma \ref{lemma:4}, and assuming $\nu^{k-1} \ge (1+c_k)/ \tilde{c}$, we get
\begin{equation}
\|\mathcal{F}_{s,1}\|_\infty\leq \|F_1^1\|_\infty+\|F_2^{1,2}\|_\infty\|(F_2^{2,2})^{-1}\|_\infty\|F_1^2\|_\infty
\le \frac{c_kN}{\nu^{k-1}}.
 \label{eq:F_1s}
\end{equation}
A similar estimate holds for $\|\tilde{\mathcal{F}}_{s,1}\|_\infty$. Hence, under similar assumptions on $\nu$
\begin{equation} \label{eq:IFs}
\begin{split}
\|I-\mathcal{F}_s\|&\leq \|I-F_0\|_\infty+\|\mathcal{F}_{s,1}\|_\infty \|\tilde{\mathcal{F}}_{s,1}\|_\infty \|\mathcal{F}_{s,2}^{-1}\|_\infty
+\|F_1^2\|_\infty \|\tilde{F}_1^2\|_\infty  \|(F_2^{2,2})^{-1}\|_\infty \\
&\le \frac{c_k}{\nu^{k-1}}+\frac{c_k}{\nu^{2(k-1)}}\frac{1}{\tilde{c}}\left( {1+ \frac{c_k}{\nu^{k-1}-c_k }} \right)
+\frac{c_k}{\nu^{2(k-1)}}\frac{1}{\tilde{c}-\frac{c_k}{\nu^{k-1}}} \\
& \le \frac{c_k}{\nu^{k-1}}.
\end{split}
\end{equation}
Moreover, 
\begin{equation}
\|\mathcal{F}_s^{-1}\|_\infty\leq \frac{1}{1-\frac{c_k}{\nu^{k-1}}}= 1+\frac{c_k}{\nu^{k-1}-c_k}. \label{eq:inv_Fs}
\end{equation}
Therefore,  for sufficiently large $\nu$, (\ref{eq:big_matrix}) is an invertible matrix. Hence, we can calculate 
the coefficient sequences by
\begin{align}
\begin{bmatrix}
\alpha \\ \beta \\ \gamma 
\end{bmatrix} =
\begin{bmatrix}
I \\ -\mathcal{F}_{s,2}^{-1}\mathcal{F}_{s,1}  \\ 
(F_2^{2,2})^{-1}(F_2^{2,1}\mathcal{F}_{s,2}^{-1}\mathcal{F}_{s,1} - F_1^2)
\end{bmatrix}
\mathcal{F}_s^{-1} u.
\end{align}
We now proceed to estimate the coefficients. We begin with the observation that
\begin{equation*}
\|\alpha\|_\infty\leq\|\mathcal{F}_s^{-1}\|_\infty \leq 1+\frac{c_k}{\nu^{k-1}-c_k}.
\end{equation*}
In addition, using (\ref{eq:invF_2s}), (\ref{eq:F_1s}) and (\ref{eq:inv_Fs}), for sufficiently large $\nu$, we get
\begin{equation*}
\begin{split}
\|\beta\|_\infty\leq& \|\mathcal{F}_{s,2}^{-1}\|_\infty\|\mathcal{F}_{s,1}\|_\infty\|\mathcal{F}_s^{-1}\|_\infty \\
&\leq \frac{c_k}{N\nu^{k-1}}.
\end{split}
\end{equation*}
Using the same estimates with additional estimates from Lemma \ref{lemma:4} give
\begin{equation*}
\begin{split}
\|\gamma\|_\infty&\leq \|(F_2^{2,2})^{-1}\| \|F_2^{1,2}\|_\infty \|\mathcal{F}_{s,2}^{-1}\|_\infty\|\mathcal{F}_{s,1}\|_\infty 
\|\mathcal{F}_s^{-1}\|_\infty \\
&\leq \frac{c_k}{N\nu^{k-1}}.
\end{split}
\end{equation*}
Finally, if $u_1=1$, we can apply (\ref{eq:IFs}), (\ref{eq:inv_Fs}) and the assumption that $|u_m|=1$, 
for each $m$, to obtain
\begin{equation*}
\begin{split}
\alpha_1&= \left( {\left( {I-(I-\mathcal{F}_s^{-1}) } \right) u} \right)_1 \\
&=u_1-\left((I-\mathcal{F}_s^{-1})u\right)_1 \\
&\geq 1-\|\mathcal{F}_s^{-1}\|_\infty\|I-\mathcal{F}_s\|_\infty \\
&\geq 1-\frac{c_k}{\nu^{k-1}}.
\end{split}
\end{equation*}
This completes the proof of Lemma \ref{lemma:1}.

\subsection{Proof of Lemma \ref{lemma:2}} \label{subsection-inversion}

Without loss of generality, assume that at $\xi_1\in\Xi$, the interpolation condition is $q(\xi_1)=1$. Let 
$\xi\in \mathbb{S}^2$ such that $d(\xi_1,{\xi})\leq \sigma/N$ for sufficiently small $0<\sigma<1$
(to be chosen later). The Hessian of $q(\xi)$ at ${\xi}$ is 
\[
H\left( q \right)\left( {\xi} \right)=\left[ 
{{\begin{array}{*{20}c}
 {\left( {D_{\xi,1} } \right)^2q\left( {\xi} \right)} 
\hfill & {D_{\xi,1} D_{\xi,2} q\left( {\xi} 
\right)} \hfill \\
 {D_{\xi,1} D_{\xi,2} q\left( {\xi} 
\right)} \hfill & {\left( {D_{\xi,2} } \right)^2q\left( 
{\xi} \right)} \hfill \\
\end{array} }} \right].
\]
We wish to show that for sufficiently small $\sigma>0$ and large enough $\nu$, $\det \left( 
{H\left( {\xi} \right)} \right)>0$ and $Tr\left( {H\left( 
{\xi} \right)} \right)<0$ , which implies that both eigenvalues 
are strictly negative and therefore $q$ is concave at $\xi$. For $r=1,2$ 
\[
\begin{array}{l}
 \left( {D_{\xi,r} } \right)^2q\left( {\xi} \right)\le 
\alpha_1 \left( {D_{\xi,r} } \right)^2F_N \left( {\xi,\xi_1 } \right)+\left\| \beta \right\|_\infty \left| {\left( 
{D_{\xi,r} } \right)^2D_{\xi_1 ,1} F_N \left( {\xi,\xi_1 } \right)} \right| \\ 
\qquad +\left\| \gamma \right\|_\infty \left| {\left( {D_{\xi,r} } 
\right)^2D_{\xi_1 ,2} F_N \left( {\xi,\xi_1 } \right)} \right| \\ 
\qquad +\left\| \alpha \right\|_\infty \sum\limits_{\xi_m \in \Xi \backslash \xi 
_1 } {\left| {\left( {D_{\xi,r} } \right)^2F_N \left( 
{\xi,\xi_m } \right)} \right|} \\ 
\qquad +\left\| \beta \right\|_\infty \left( {\sum\limits_{\xi_m \in \Xi 
\backslash \xi_1 } {\left| {\left( {D_{\xi,r} } \right)^2D_{\xi 
_m ,1} F_N \left( {\xi,\xi_m } \right)} \right|} } \right) \\ 
\qquad +\left\| \gamma \right\|_\infty \left( {\sum\limits_{\xi_m \in \Xi 
\backslash \xi_1 } {\left| {\left( {D_{\xi,r} } \right)^2D_{\xi 
_m ,2} F_N \left( {\xi,\xi_m } \right)} \right|} } \right). \\ 
 \end{array}
\]
We estimate the first left hand term using (\ref{eq:alpha0}), (\ref{2-ord-der-diag}), (\ref{bound-alpha}) and 
then (\ref{eq:lip-3})
\[
\begin{split}
 \alpha_1 \left( {D_{\xi,r} } \right)^2F_N \left( {\xi,\xi_1 } \right)&=\alpha_1 \left( {D_{\xi,r} } \right)^2F_N 
\left( {\xi,\xi} \right)+\alpha_1 \left( {\left( 
{D_{\xi,r} } \right)^2F_N \left( {\xi,\xi_1 } 
\right)-\left( {D_{\xi,r} } \right)^2F_N \left( {\xi,\xi } \right)} \right) \\ 
& \le -\left( {1-\frac{c_k }{\nu^{k-1}}} \right){F}'_N \left( 1 
\right)+\left( {1+\frac{c_k }{\nu^{k-1}-c_k }} \right)c_k N^3d\left( 
{\xi,\xi_1 } \right) \\ 
& \le -N^2\left( {\tilde{c} \left( {1-\frac{c_k }{\nu^{k-1}}} \right)-\left( 
{1+\frac{c_k }{\nu^{k-1}-c_k }} \right)c_k \sigma } \right). \\ 
 \end{split}
\]
The next two terms are estimated using the bounds on $\alpha,\beta$ (\ref{bound-beta}), 
(\ref{bound-gamma}) and (\ref{eq:third-der})
\[
\left\| \beta \right\|_\infty \left| {\left( {D_{\xi,r} } 
\right)^2D_{\xi_1 ,1} F_N \left( {\xi,\xi_1 } \right)} 
\right|,\left\| \gamma \right\|_\infty \left| {\left( {D_{\xi,r} } 
\right)^2D_{\xi_1 ,2} F_N \left( {\xi,\xi_1 } \right) } \right| \le 
\frac{c_k }{\nu^{k-1}}N^2.
\]
Estimates (\ref{bound-alpha}) and (\ref{point-kernel-estimate_2}) give
\[
\left\| \alpha \right\|_\infty \sum\limits_{\xi_m \in \Xi \backslash \xi_1 
} {\left| {\left( {D_{\xi,r} } \right)^2F_N \left( {\xi,\xi_m } \right)} \right|} \le \left( {1+\frac{c_k }{\nu^{k-1}-c_k }} 
\right)\frac{c_k }{\nu^{k-1}}N^2.
\]
Using (\ref{bound-beta}), (\ref{bound-gamma}) and (\ref{point-kernel-estimate_3})
\[
\left\| \beta \right\|_\infty \left( {\sum\limits_{\xi_m \in \Xi \backslash 
\xi_1 } {\left| {\left( {D_{\xi,r} } \right)^2D_{\xi_1 ,1} F_N 
\left( {\xi,\xi_1 } \right)} \right|} } \right),\left\| \gamma 
\right\|_\infty \left( {\sum\limits_{\xi_m \in \Xi \backslash \xi_1 } 
{\left| {\left( {D_{\xi,r} } \right)^2D_{\xi_1 ,2} F_N \left( 
{\xi,\xi_1 } \right)} \right|} } \right)\le \frac{c_k }{\nu 
^{k-1}}N^2.
\]
Thus, for sufficiently small $\sigma $ and large $\nu $ 
\[
\left( {D_{\xi,r} } \right)^2q\left( {\xi} \right)\le 
-N^2\left( {\tilde{c} \left( {1-\frac{c_k }{\nu^{k-1}}} \right)-\left( 
{1+\frac{c_k }{\nu^{k-1}-c_k }} \right)c_k \sigma +\frac{c_k }{\nu^{k-1}}} 
\right)<0.
\]
We proceed with the estimate of the two other entries of the Hessian
\[
\begin{array}{l}
 \left| {D_{\xi,1} D_{\xi,2} q\left( {\xi} 
\right)} \right|\le \alpha_1 \left| {D_{\xi,1} D_{\xi,2} F_N \left( {\xi,\xi_1 } \right)} \right|+\left\| \beta 
\right\|_\infty \left| {D_{\xi,1} D_{\xi,2} D_{\xi_1 
,1} F_N \left( {\xi,\xi_1 } \right)} \right| \\ 
 +\left\| \gamma \right\|_\infty D_{\xi,1} D_{\xi,2} 
D_{\xi_1 ,2} F_N \left( {\xi,\xi_1 } \right) \\ 
 +\left\| \alpha \right\|_\infty \sum\limits_{\xi_m \in \Xi \backslash \xi 
_1 } {\left| {D_{\xi,1} D_{\xi,2} F_N \left( 
{\xi,\xi_m } \right)} \right|} \\ 
 +\left\| \beta \right\|_\infty \left( {\sum\limits_{\xi_m \in \Xi 
\backslash \xi_1 } {\left| {D_{\xi,1} D_{\xi,2} 
D_{\xi_1 ,1} F_N \left( {\xi,\xi_1 } \right)} \right|} } 
\right) \\ 
 +\left\| \gamma \right\|_\infty \left( {\sum\limits_{\xi_m \in \Xi 
\backslash \xi_1 } {\left| {D_{\xi,1} D_{\xi,2} 
D_{\xi_1 ,2} F_N \left( {\xi,\xi_1 } \right)} \right|} } 
\right). \\ 
 \end{array}
\]
Using first (\ref{bound-alpha}), (\ref{mix-part-zero}) and then (\ref{eq:lip-3}) yields
\[
\begin{split}
 \alpha_1 \left| {D_{\xi,1} D_{\xi,2} F_N \left( 
{\xi,\xi_1 } \right)} \right| &\le \left( {1+\frac{c_k }{\nu 
^{k-1}-c_k }} \right)\left| {D_{\xi,1} D_{\xi,2} F_N 
\left( {\xi,\xi_1 } \right)-D_{\xi,1} D_{\xi,2} F_N \left( {\xi,\xi} \right)} \right| \\ 
&\le \left( {1+\frac{c_k }{\nu^{k-1}-c_k }} \right)c_k N^3d\left( 
{\xi,\xi_1 } \right) \\ 
&\le \left( {1+\frac{c_k }{\nu^{k-1}-c_k }} \right)c_k \sigma N^2. \\ 
 \end{split}
\]
Combining with similar estimates as in the previous case results in
\[
\left| {D_{\xi,1} D_{\xi,2} q\left( {\xi} 
\right)} \right|\le N^2\left( {\left( {1+\frac{c_k }{\nu^{k-1}-c_k }} 
\right)c_k \sigma +\frac{c_k }{\nu^{k-1}}+\left( {1+\frac{c_k }{\nu 
^{k-1}-c_k }} \right)\frac{c_k }{\nu^{k-1}}} \right).
\]
It is now clear, that we can chose sufficiently small $\sigma $ and 
large enough $\nu$ such that 
$\left| {D_{\xi,1} D_{\xi,2} q\left( {\xi} 
\right)} \right|<\left| {\left( {D_{\xi,r} } \right)^2q\left( 
{\xi} \right)} \right|$ and $\left( {D_{\xi,r} } 
\right)^2q\left( {\xi} \right)<0$, $r=1,2$. This gives that $\det \left( 
{H\left( {\xi} \right)} \right)>0$ and $Tr\left( {H\left( 
{\xi} \right)} \right)<0$.
To finish the proof, we have to show that $q(\xi)>-1$
\[
\begin{split}
q(\xi)&\geq \alpha_0F_N(\xi\cdot\xi_1)-\|\alpha\|_\infty\sum_{\xi_m\in\Xi\backslash\xi_1}\vert F_N(\xi\cdot\xi_m)\vert \\
& -\|\beta\|_\infty\sum_{\xi_m\in\Xi}\vert D_{\xi ,1 }F_N(\xi,\xi_m)\vert -\|\gamma\|_\infty\sum_{\xi_m\in\Xi}\vert D_{\xi ,2 }F_N(\xi,\xi_m)\vert \\
&\geq \left(1-\frac{c}{\nu^{k-1}}\right)(1+F_N(\xi\cdot\xi_1)-F_N(\xi\cdot\xi))-\left(1+\frac{c_k}{\nu^{k-1}}\right)\frac{c_k}{\nu^{k-1}}-\frac{2c_k}{\nu^{2(k-1)}} \\
&\geq \left(1-\frac{c}{\nu^{k-1}}\right)(1-c_k \sigma)-\frac{2c_k}{\nu^{2(k-1)}}.
\end{split}
\]
Clearly, for large $\nu$ and small $\sigma$, $q(\xi)>-1$. For the case where $q(\xi_1)=-1$, the proof is almost identical
except for the fact that we show that $q$ is convex in the neighborhood of $\xi_1$ and $q(\xi)<1$, 
for $d(\xi,\xi_1)<\sigma / N$.

\subsection{Proof of Lemma \ref{lemma:3}}

Let $\xi \in \mathbb{S}^2$ and $\xi_1 \in \Xi$, such that $\sigma/N \le d(\xi,\xi_1) \le \Delta/2$. We need to show 
that for sufficiently large $\nu$, $|q(\xi)| < 1$. First observe that using only the first order estimate
for $F_N(\xi \cdot \xi_1)$, with the normalization $F_N(\xi_1,\xi_1)=1$ 

\[
\left| {\alpha_1 } \right|\left| {F_N \left( {\xi \cdot \xi_1 } \right)} 
\right|\le \left( {1+\frac{c_k }{\nu^{k-1}}} \right)\frac{1}{1+\sigma }.
\]
Consequently, for sufficiently large $\nu$, using also the estimates of Lemmas \ref{lemma:1},
 \ref{lemma-sum-estimate} and (\ref{eq:estimate-diff-1}) gives

\[
\begin{split}
 \left| {q\left( \xi \right)} \right|&\le \left\| \alpha \right\|_\infty 
\left| {F_N \left( {\xi \cdot \xi_1 } \right)} \right|+\left\| \beta 
\right\|_\infty \left| {D_{\xi_1 ,1} F_N \left( {\xi ,\xi_1 } \right)} 
\right|+\left\| \gamma \right\|_\infty \left| {D_{\xi_1 ,2} F_N \left( {\xi 
,\xi_1 } \right)} \right| \\ 
& \qquad +\left\| \alpha \right\|_\infty \sum\limits_{\xi_m \in \Xi \backslash \xi 
_1 } {\left| {F_N \left( {\xi \cdot \xi_m } \right)} \right|} +\left\| 
\beta \right\|_\infty \sum\limits_{\xi_m \in \Xi \backslash \xi_1 } 
{\left| {D_{\xi_1 ,1} F_N \left( {\xi ,\xi_m } \right)} \right|} \\
&\qquad + \left\| 
\gamma \right\|_\infty \sum\limits_{\xi_m \in \Xi \backslash \xi_1 } 
{\left| {D_{\xi_1 ,2} F_N \left( {\xi ,\xi_m } \right)} \right|} \\ 
& \le \left( {1+\frac{c_k }{\nu^{k-1}}} \right)\frac{1}{1+\sigma 
}+\frac{2c_k }{\nu^{k-1}}\frac{c_k }{\left( {1+\sigma } \right)^k}+\left( 
{1+\frac{c_k }{\nu^{k-1}}} \right)\frac{c_k }{\nu^{k-1}}+\frac{2c_k }{\nu 
^{2\left( {k-1} \right)}} \\ 
& <1. \\ 
 \end{split}
\]
The case where $d(\xi,\xi_m)>\Delta/2$, for each $\xi_m \in \Xi$ is easier. In this case, where $\xi$ is well separated 
from all the points of $\Xi$, we can use estimates similar to the those of Lemma \ref{lemma-sum-estimate}, to get
\[
\begin{split}
\vert q(\xi)\vert&\leq \|\alpha\|_\infty\sum_{\xi_m\in\Xi}\vert F_N(\xi\cdot\xi_m)\vert +\|\beta\|_\infty\sum_{\xi_m\in\Xi}\vert D_{\xi_m ,1 }F_N(\xi\cdot\xi_m)\vert \\
&\qquad +\|\gamma\|_\infty\sum_{\xi_m\in\Xi}\vert D_{\xi_m,2 }F(\xi\cdot\xi_m)\vert \\
&\leq \left(1+\frac{c_k}{\nu^{k-1}}\right)\frac{c_k}{\nu^{k-1}} +\frac{c_k}{\nu^{2(k-1)}}.
\end{split}
\]

This concludes the proof. 

\section{Non-Negative Signals}
\label{sec:non-negative}
In this section, we show that for the special case of non-negative Dirac ensembles 
\begin{equation}
f=\sum_mc_m\delta_{\xi_m} \,,\, c_m>0\,,\,\xi_m\in\Xi, \label{eq:non_negative_signal}
\end{equation}
a sparsity condition is sufficient for exact recovery (compare with the discrete case \cite{donoho-tanner}).
We start by presenting a sufficient condition for the reconstruction of the signal from its projection onto 
$V_N$. Here we give a general version of the theorem as follows:
\begin{theorem}
Let $f=\sum_mc_m\delta_{\xi_m}$, where $\Xi=\{\xi_m\}\subset A$, with $A$ a compact manifold in $\mathbb{R}^d$ and $c_m>0$. 
Let $\Pi_D$ be a linear space of continuous functions of dimension $D$ in $A$. For any basis $\{P_k\}_{k=1}^D$ of $\Pi_D$,
let $y_k=\langle f,P_k\rangle$ for all $1\leq k\leq D$. If there exists $q\in \Pi_D$ such that
\begin{align}
q(\xi_m)&=1 \quad \xi_m\in  \Xi , \label{eq:con_nng_1}  \\
\vert q(\xi)\vert& <1 \quad \xi\notin  \Xi, \label{eq:con_nng_2} 
\end{align}
then, $f$ is the unique minimizer over all non-negative measures of the following 
\begin{equation}
min_{g \in\mathcal{M(A)}}\|g\|_{TV} \quad s.t. \quad y_k=\langle g,P_k\rangle , k=1,\dots,D.
\label{eq:TV}
\end{equation}

\begin{proof}

Let $g$ be the solution of (\ref{eq:TV}), and set $g=f+h, h\neq 0$. Let $h=h_\Xi+h_{\Xi^C}$ be the Lebesgue decomposition of $h$ relative to $\vert f\vert$, so that $h_\Xi$ is supported on $\Xi$. Additionally, $h_\Xi=\sum d_m\delta_{\xi_m}$ for some real $\{d_m\}$. Also, since $g$ is a non-negative measure, $f+h_\Xi$ is also non-negative, implying $c_m+d_m\geq 0$  for all $\xi_m\in \Xi$. Thus, $\|f+h_\Xi\|_{TV}=\sum_m\left(c_m+d_m\right)$.

We observe that 
\begin{equation}
0=\langle q,h\rangle=\langle q,h_\Xi\rangle+\langle q,h_{\Xi^C}\rangle=\sum_m d_m+\langle q,h_{\Xi^C}\rangle.
\end{equation}

Plainly, if $h_{\Xi^C}= 0$, then $h_\Xi=0$, and consequently $h=0$.
Else, if $h_{\Xi^C}\neq 0$, we obtain
\begin{equation}
\left\vert \sum_md_m\right\vert =\left\vert \int qdh_{\Xi^C}\right\vert <\|h_{\Xi^C}\|_{TV}. 
\end{equation}
This leads to the following contradiction
\begin{equation}
\begin{split}
\|f\|_{TV}&\geq \|f+h\|_{TV}=\|f+h_\Xi\|_{TV}+\|h_{\Xi^C}\|_{TV} \\
&>\sum_m\left(c_m+d_m\right)+\left\vert \sum_md_m\right\vert \\
&=\|f\|_{TV}+\left\vert \sum_md_m\right\vert+ \sum_md_m\geq \|f\|_{TV}
\end{split}
\end{equation}

Therefore, $f=g$.
\label{th:non_negative}
\end{proof}
\end{theorem}

We now show that a polynomial $q\in V_N(\mathbb{S}^{d-1})$, $d \ge 2$, obeying (\ref{eq:con_nng_1}) and (\ref{eq:con_nng_2}) 
can be constructed with a sparsity condition replacing the separation condition. Assuming that 
$\vert\Xi\vert=s\leq N$, we construct the following polynomial 
\begin{equation}
q(\xi):=1-2^{-(s+1)}\prod_{m=1}^s\left(1-\xi\cdot\xi_m\right). \label{eq:non_negative_poly}
\end{equation}
As already noted, the function $G(\xi)= \xi \cdot \xi_0$ is a spherical harmonic and thus also $1-G(\xi)$. 
The fact that a product of spherical harmonics of degrees $N_1,N_2$ is a spherical harmonic of 
degree $N_1+N_2$ and the computation of the corresponding representation is known as Clebsch - Gordan. 
Plainly, as long as $s\leq N$, $q\in V_N$. Moreover, $q(\xi_m)=1$, and $0 \le q(\xi)<1$ for any $\xi\notin\Xi$.

As a result of the above construction, we may apply Theorem \ref{th:non_negative} to obtain exact recovery for non-negative Dirac ensembles whenever the sparsity condition $\vert\Xi\vert\leq N$ holds.

Observe that the case of univariate non-negative Dirac trains and spaces of trigonometric polynomials is a special case of the above, with $d=2$. Therefore, a sparsity condition can replace the separation condition of \cite{candes2013towards}.
For $d=2$, the construction of the interpolating polynomial over knots $\{ t_m \} \subset [-\pi,\pi]$, takes the form
\begin{equation}
q(t)=1-2^{-(s+1)}\prod_{j=1}^s\left(1-\cos(t-t_m)\right), \qquad t \in [-\pi,\pi].
\end{equation}

Similarly, in \cite{bendory2013exact} the authors showed that the separation condition is a sufficient condition for the reconstruction of signals of the form (\ref{eq:signal}) from their projection onto the space of algebraic polynomials of degree $N$ over $[-1,1]$. If the signal is known to be non-negative, a sufficient condition for reconstruction is  $\vert\Xi\vert\leq N/2$, by the construction of the following algebraic polynomial 
(see also \cite{de2012exact})
\begin{equation}
q(\xi)=1-4^{-(s+1)}\prod_{i=1}^s(\xi-\xi_m)^2.
\end{equation}


\end{document}